\documentclass[twocolumn,prb]{revtex4-1}
\usepackage{graphics}

\usepackage{graphics}
\usepackage{enumerate}
\usepackage{soul}
\usepackage{color}
\newcommand{\beq}{\begin{equation}}
\newcommand{\eeq}{\end{equation}}
\newcommand{\beqa}{\begin{eqnarray}}
\newcommand{\eeqa}{\end{eqnarray}}

\begin{document}

\title{Minimum Uncertainty and Entanglement }

\author{N.D. Hari Dass}
\address{Chennai Mathematical Institute, Chennai, India.\\
CQIQC, Indian Institute of Science, Bangalore, India.\\
dass@cmi.ac.in}
\author{Tabish Qureshi}
\address{Centre for Theoretical Physics, Jamia Millia Islamia,
New Delhi, India.\\
tabish@ctp-jamia.res.in}
\author{Aditi Sheel}
\address{Department of Physics, Jamia Millia Islamia, New Delhi, India.\\
mail2adt@gmail.com}


\begin{abstract}
We address the question, does a system A being entangled with another system B, 
put any constraints on the Heisenberg uncertainty relation (or the
Schr\"odinger-Robertson inequality)?
We find that the equality of the uncertainty relation cannot be reached
for any two noncommuting observables, for finite dimensional Hilbert spaces 
if the Schmidt rank of the entangled state is \emph{maximal}.
One consequence is that the lower bound of the uncertainty relation can never be
attained for any two observables for {\em qubits}, if the state is entangled.
For infinite-dimensional Hilbert space too, we show that there is a class
of physically interesting entangled states for which no two noncommuting 
observables can attain the minimum uncertainty equality.
\end{abstract}

\maketitle
\section{Introduction}

In quantum mechanics, the product of uncertainties of two noncommuting
observables is bounded.
It must respect
the Heisenberg Uncertainty Relation (HUR)\cite{heisenberg}
\begin{equation}
(\Delta X)^2(\Delta Y)^2 \ge {1\over 4}|\langle[{\mathbf X},{\mathbf Y}]\rangle|^2 ,
\label{hur}
\end{equation}
Another form of uncertainty relation, 
the Schr\"{o}dinger-Robertson
inequality (SR),\cite{robertson}
\begin{equation}
\label{sr}
(\Delta X)^2(\Delta Y)^2 \ge {1\over 4}|\langle[{\mathbf X},{\mathbf Y}]\rangle|^2 
+{1\over 4}|\langle \{{\tilde{\mathbf X}},{\tilde{\mathbf Y}} \}\rangle|^2
\end{equation}
where, ${\tilde{\mathbf O}}={\mathbf O} -\langle {\mathbf O} \rangle$,
has also been studied. In general SR provides a stronger bound on the 
uncertainties, as compared to HUR, because the r.h.s. of (\ref{sr})
is greater than or equal to the r.h.s. of (\ref{hur}).
Entanglement is another fundamental aspect of quantum mechanics.
An interesting
study of SR for entangled systems has been worked out in Ref. \onlinecite{nha}.
The interplay between 
entanglement and the uncertainty relation
has been explored in a number of interesting papers.
\cite{hofmann,giovannetti,guehne,lewenstein,guhne,rajgopal,nha7,wang,anastopoulos,zhang}

In this paper, we ask the following question - if a system is entangled
with another system, does this entanglement have any bearing
on the uncertainty between two observables {\em of the same system}?

We proceed by examining explicit
examples to show that it does, that for many cases, the equality can
never be achieved
for entangled states. We then go on to prove, on rather general grounds,
that the equality cannot be attained in many cases.
\section{Specific Examples}

We start with the simplest bipartite entangled state we can think of, consisting of
only two parts
\beq
|\Psi\rangle = c_1|\psi_1\rangle_A|\alpha_1\rangle_B + c_2|\psi_2\rangle_A|\alpha_2\rangle_B 
\label{entangled}
\eeq
where $|\psi_i\rangle_A$  are two states of system A
, and $|\alpha_j\rangle_B$ are two {\em orthonormal}
states of system B. The constants $c_1, c_2$ satisfy
$|c_1|^2+|c_2|^2=1$.
The uncertainties in two observables ${\mathbf X}_A\otimes {\mathbf 1}_B$ and ${\mathbf Y}_A\otimes {\mathbf 1}_B$,
in the entangled state $|\Psi\rangle$, are defined as
\begin{equation}
(\Delta X)^2_\Psi = \langle\Psi|{\tilde{\mathbf X}^2}|\Psi\rangle \quad\quad
 (\Delta Y)^2_\Psi = \langle\Psi|{\tilde{\mathbf Y}^2}|\Psi\rangle 
\end{equation}
In what follows, we suppress the direct products explicitly as all our observables will be operating
on system A. We also use the shorthand $\langle {\mathbf O} \rangle_\Psi$ for $\langle \Psi|{\mathbf O}|\Psi\rangle$,
and $\langle {\mathbf O}_A \rangle_i$ for $\langle \psi_i|{\mathbf O}|\psi_i\rangle_A$.

One can relate the uncertainties in $|\Psi\rangle$ to the uncertainties of the observables
in the states $|\psi_i\rangle$.
For a generic observable ${\mathbf O}$ 
\begin{equation}
(\Delta O)^2_\Psi = \sum_i~|c_i|^2(\Delta O)_i^2 + 
                 |c_1|^2|c_2|^2(\langle {\mathbf O}\rangle_1 - \langle {\mathbf O}\rangle_2)^2 
\label{deltaab}
\end{equation}
where
$(\Delta O)_i^2 = \langle\psi_i|{\tilde{\mathbf O}}^2|\psi_i\rangle$. 
The product of uncertainties can be worked out to be 
\begin{eqnarray}
(\Delta X)^2_\Psi(\Delta Y)^2_\Psi &=& |c_1|^4(\Delta X)_1^2(\Delta Y)_1^2
+|c_2|^4(\Delta X)_2^2(\Delta Y)_2^2
\nonumber\\
&&+ 2|c_1|^2|c_2|^2(\Delta X)_1(\Delta Y)_1
(\Delta X)_2(\Delta Y)_2\nonumber\\
&&+ |c_1|^4|c_2|^4(\langle {\mathbf Y}\rangle_1 - \langle {\mathbf Y}\rangle_2)^2 
(\langle {\mathbf X}\rangle_1 - \langle {\mathbf X}\rangle_2)^2\nonumber\\ 
&&+ |c_1|^2|c_2|^2 
\{(\Delta X)_1(\Delta Y)_2-(\Delta X)_2(\Delta Y)_1\}^2\nonumber\\
&&+|c_1|^2|c_2|^2\{\sum_i~|c_i|^2(\Delta X)_i^2
\cdot
(\langle {\mathbf Y}\rangle_1
- \langle {\mathbf Y}\rangle_2)^2\nonumber\\
&&+\sum_i~|c_i|^2(\Delta Y)_i^2\cdot
(\langle {\mathbf X}\rangle_1 - \langle {\mathbf X}\rangle_2)^2\} 
\label{lhs}
\end{eqnarray}

It follows from (\ref{lhs}) that the necessary conditions for the l.h.s. to reach its minimum value
are (i) $\langle {\mathbf X} \rangle_1 = \langle {\mathbf X} \rangle_2$, (ii) $\langle {\mathbf Y} \rangle_1 =
\langle {\mathbf Y} \rangle_2$, (iii) the states $|\psi\rangle_i$ be minimum uncertainty states themselves, and (iv) $(\Delta X)_1(\Delta Y)_2=
(\Delta Y)_1(\Delta X)_2$. In the examples that follow we show that
entangled states do not saturate the equality in HUR, for a wide range of
familiar systems. 

{\bf (a) Angular Momentum Operators:}
The HUR between, say,
${\mathbf J}_x$ and ${\mathbf J}_y$ reads
\begin{equation}
(\Delta J_x)^2(\Delta J_y)^2 \ge {1\over 4}|\langle i\hbar {\mathbf J}_z\rangle|^2
\end{equation}
Consider the state 
\begin{equation}
|\Psi\rangle = c_1|m_1\rangle|\alpha_1\rangle + c_2|m_2\rangle|\alpha_2\rangle ,
\end{equation}
where $|m_1\rangle,|m_2\rangle$ are two of the
eigenstates of ${\mathbf J}_z$.
The uncertainties for $|m\rangle$ are given by
$(\Delta J_x)^2 = (\Delta J_y)^2 = {\hbar^2\over 2}(j(j+1)-m^2)$,
which will be minimum for $m=\pm j$. 
So $|\Psi\rangle$ can be entangled
only if $m_1=+j$ and $m_2=-j$, or vice-versa. 
As the expectation values of both $J_x,J_y$ in eigenstates of $J_z$ are zero, this example satisfies
all the conditions i-iv. 
The necessary conditions do not further
restrict $|\psi\rangle$. 

But, $(\Delta J_x)^2_\Psi\cdot(\Delta J_y)^2_\Psi = \frac{j^2\hbar^2}{4}$
whereas $|\langle {\mathbf J}_z \rangle_{\Psi}|^2 = j^2\hbar^2(|c_1|^2-|c_2|^2)^2$. 
Therefore, there can be equality in HUR
only when one of the $c_i$ vanishes, but then $|\Psi\rangle$ is not entangled.
Thus we conclude that 
\begin{equation}
(\Delta J_x)^2(\Delta J_y)^2 > {1\over 4}|\langle i\hbar {\mathbf J}_z\rangle|^2 
\end{equation}
for entangled states of a system with fixed angular momentum.


{\bf (b) Heisenberg Algebra:} Next we look at the position and momentum operators ,
${\mathbf X}$ and ${\mathbf P}$ in one dimension. The HUR has the form
\begin{equation}
(\Delta X)^2(\Delta P)^2 \ge {\hbar^2\over 4} .
\end{equation}
We consider an entangled state made up of
two Gaussian states entangled with two orthogonal states of another system.
The Gaussian states are described by 
\begin{equation}
\langle x|\psi_i\rangle = {1\over(2\pi\sigma_i^2)^{1/4}} e^{ip_ix/\hbar}
\exp\left(-{(x-x_i)^2\over 4\sigma_i^2}\right) 
\end{equation}
Hence in this case $\langle{\mathbf X}\rangle_i = x_i$, 
$\langle{\mathbf P}\rangle_i = p_i$, 
$(\Delta X)_i = \sigma_i$, 
$(\Delta p)_i = {\hbar\over 2\sigma_i}$. 

The conditions i-iv yield: 
$x_1=x_2$, $p_1=p_2$ and $\sigma_1=\sigma_2$. But in that situation, both
the Gaussians are identical, and hence the state saturating the equality
in HUR is disentangled. 

As yet another example consider an entangled state built out
of energy-eigenstates of Harmonic oscillator,
\begin{equation}
|\Psi\rangle = c_1|n_1\rangle|\alpha_1\rangle + c_2|n_2\rangle|\alpha_2\rangle ,
\label{sho}
\end{equation}
where the states $|n_i\rangle$ 
satisfy ${\mathbf H}|n_i\rangle=(n_i+{1\over 2})
\hbar\omega|n_i\rangle$, where ${\mathbf H}$ is the
Hamiltonian of a Harmonic oscillator with frequency
$\omega$. Now,
$\langle n|{\mathbf X}|n\rangle = \langle n|{\mathbf P}|n\rangle = 0$
for any $|n\rangle$. 
Also, the uncertainties can be easily calculated to yield,
$(\Delta X)_n^2=(2n+1){\hbar\over 2m\omega}$ and
$(\Delta P)_n^2=(2n+1){\hbar m\omega\over 2}$. In this case too, 
all the conditions i-iv are satisfied.

Nevertheless, equation (\ref{lhs}) then assumes the form
\begin{equation}
(\Delta X)^2_\Psi(\Delta P)^2_\Psi =
{\hbar^2\over 4}\sum_i |c_i|^2(2n_i+1) 
\end{equation}
$(\Delta X)^2(\Delta P)^2$ can be equal to $\hbar^2/4$ {\em only if}
$n_1=n_2=0$. But in that case, the state (\ref{sho}) becomes
disentangled. 
Again we see that minimum uncertainty equality cannot be
achieved as long as the state is entangled.

Lastly we consider the example of a 
continuous variable entangled state which is a superposition of an
infinite number of parts.
\begin{equation}
\Psi(x_A,x_B) = {1\over \sqrt{\pi\Omega/\sigma}}
 e^{-(x_A-x_B)^2\sigma^2}
e^{-(x_A+x_B)^2/16\Omega^2} 
\label{newstate}
\end{equation}
In the limit $\sigma\to\infty$ and
$\Omega\to\infty$, the state reduces to the so-called EPR state considered
by Einstein, Podolsky and Rosen.\cite{epr}
The uncertainties in position and momentum of particle A (say) is given by
\begin{equation}
\Delta X_A = \sqrt{\Omega^2+1/16\sigma^2},~~~
\Delta P_{A} = \hbar\sqrt{\sigma^2 + {1\over 16\Omega^2}}. \label{dp}
\end{equation}
The minimum uncertainty equality is obtained only if $\Omega={1\over{4\sigma}}$.
But for these values, the state becomes disentangled, as one can see from
(\ref{newstate}). 

Based on these examples, we had initially conjectured that the lower bound
of the HUR and the SR cannot be obtained for any two observables if the
state is entangled. However, Englert provides counter-examples to show
situations where lower bound is obtained for entangled states.\cite{englert}

In the following we will carry out a general analysis, and prove that there
is a wide class of scenarios in which this lower bound cannot be achieved
by entangled states.

\section{General analysis}
\subsection{Finite Dimensional Hilbert Spaces}
Let $|\Psi\rangle$ be a \emph{pure} entangled state of two quantum systems belonging to Hilbert spaces $({\cal H}_A,{\cal H}_B)$,
with respective dimensionalities $d_A,d_B$ and let $d_A\le d_B$. The entangled state
$|\Psi\rangle$ admits a Schmidt decomposition
\begin{equation}
\label{schmidt}
|\Psi\rangle = \sum_i^s~c_i|a_i\rangle_A~|b_i\rangle_B
\end{equation}
where $|a_i\rangle_A,|b_i\rangle_B$ are orthonormal basis vectors in ${\cal H}_A,{\cal H}_B$ respectively. 
The number $s\le d_A$ is called the Schmidt rank. \\

Now let us consider single-system observables acting
on ${\cal H}_A$. Treatment of when the observables act on ${\cal H}_B$ is completely parallel. 
We are considering operators of the type ${\mathbf O}_A\otimes {\bf 1}_B$. Consider
a pair of such Hermitian operators $ {\mathbf X}_A,  {\mathbf Y}_A$ that do not commute with each other, i.e 
$[{\mathbf X}_A,{\mathbf Y}_A] = {\mathbf C}_A \ne 0$.

Schwarz inequality for the states ${\mathbf X}_A|\Psi\rangle,{\mathbf Y}_A|\Psi\rangle$ gives
\begin{equation}
\label{schwarz}
\langle  {\mathbf X}_A^2\rangle_\Psi\langle {\mathbf Y}_A^2\rangle_\Psi
\ge |\langle({\mathbf X}_A {\mathbf Y}_A)\rangle_\Psi|^2
\end{equation}
The inner product occurring on the r.h.s. of (\ref{schwarz}) can be written as
\begin{equation}
\label{schwarzrhs}
\langle {\mathbf X}_A {\mathbf Y}_A\rangle_\Psi = \frac{1}{2}\langle \{{\mathbf X}_A, {\mathbf Y}_A\}
\rangle_\Psi+\frac{1}{2}\langle [{\mathbf X}_A,{\mathbf Y}_A]\rangle_\Psi
\end{equation}
The first term is purely real while the second term is purely imaginary. Hence (\ref{schwarz})
can be rewritten as
\begin{equation}
\langle  {\mathbf X}_A^2\rangle_\Psi\langle {\mathbf Y}_A^2\rangle_\Psi
 \ge \frac{1}{4}|\langle {\mathbf C}_A\rangle_\Psi|^2
+\frac{1}{4}|\langle\{{\mathbf X}_A,{\mathbf Y}_A\}\rangle_\Psi|^2
\end{equation}
We now consider the operators ${\mathbf{\tilde X}}_A={\mathbf X}_A-\langle {\mathbf X}_A\rangle_\Psi,
{\mathbf{\tilde Y}_A}={\mathbf Y}_A-\langle {\mathbf Y}_A\rangle_\Psi$ 
instead of the operators ${\mathbf X}_A,{\mathbf Y}_A$ 
respectively.
Then we can put together everything and write
\begin{equation}
\label{tildeschwarz}
\langle{\mathbf {\tilde X}}_A^2\rangle_\Psi
\cdot \langle{\mathbf {\tilde Y}}_A^2\rangle_\Psi
\ge \frac{1}{4}|\langle[{\mathbf X}_A,{\mathbf Y}_A]\rangle_\Psi|^2
+\frac{1}{4} |\langle\{{\mathbf {\tilde X}}_A,{\mathbf {\tilde Y}}_A\} 
\rangle_\Psi|^2\nonumber\\
\end{equation}
The l.h.s. of (\ref{tildeschwarz}) is the same as
$(\Delta X_A)_\Psi^2\cdot(\Delta Y_A)_\Psi^2$.
Thus, (\ref{tildeschwarz}) is nothing but the SR inequality.
The equality in (\ref{schwarz}) holds if, and only if, the vectors 
$ {\mathbf X}_A|\Psi\rangle,  {\mathbf Y}_A|\Psi\rangle$ are parallel. That is, if there exists 
a complex number $\gamma$ such
that
\begin{equation}
\label{parallel}
 {\mathbf X}_A|\Psi\rangle+\gamma~ {\mathbf Y}_A|\Psi\rangle = 0
\end{equation}
This, in addition to leading to the equality in eqn.(\ref{schwarz}), further implies that
\begin{eqnarray}
\label{extraschwarz}
&&\gamma_R~\langle [{\mathbf X}_A,{\mathbf Y}_A]\rangle_\Psi+i\gamma_I~\langle 
\{{\mathbf X}_A,{\mathbf Y}_A\}\rangle_\Psi=0\nonumber\\
&&(\Delta X_A)^2 = |\gamma|^2~(\Delta Y_A)^2
\end{eqnarray}
Therefore, for the equality in HUR to be realised, the last term in (\ref{tildeschwarz}), which is real,
must also vanish in addition eqn.(\ref{parallel}), but now for the 
new set of operators ${\mathbf {\tilde X}}_A,{\mathbf {\tilde Y}}_A$:
\begin{equation}
\label{parallel2}
 ({\mathbf X}_A-\langle {\mathbf X}_A \rangle_\Psi)|\Psi\rangle+\Gamma~ ({\mathbf Y}_A-\langle {\mathbf Y}_A \rangle_\Psi)|\Psi\rangle = 0
\end{equation}

 This is possible only if
$\Gamma$ appearing in (\ref{parallel2}) is {\em purely imaginary}. For the SR case, however, $\Gamma$ can
be \emph{any} complex number.
Substituting (\ref{schmidt}) in (\ref{parallel2}):
\begin{equation}
\label{parallel3}
\sum_{i=1}^s~c_i~\{ {\mathbf {\tilde X}}_A+\Gamma~ {\mathbf {\tilde Y}_A\}}
|a_i\rangle_A|b_i\rangle_B = 0
\end{equation}
This can only be satisfied if
\begin{equation}
\label{subparallel}
\{({\mathbf { X}}_A-\langle {\mathbf X}_A\rangle_\Psi)+\Gamma~({\mathbf { Y}}_A-\langle {\mathbf Y}_A\rangle_\Psi)\}
|a_i\rangle_A = 0
\end{equation}
for every i. Taking the inner product of this equation with $|a_i\rangle_A$, one gets:
\begin{equation}
(\langle {\mathbf X}_A \rangle_i - \langle {\mathbf X}_A \rangle_\Psi) +\Gamma~(\langle {\mathbf Y}_A \rangle_i-\langle {\mathbf Y}_A \rangle_\Psi)=0
\end{equation}
for every i. Here $\langle {\mathbf O}_A \rangle_i$ is the expectation value of ${\mathbf O}_A$ in $|a_i\rangle_A$.
But due to the 
real nature of all the expectation values, this is possible if and only if
\begin{equation}
\langle {\mathbf X}_A \rangle _i = \langle {\mathbf X}_A \rangle_\Psi~~\quad\quad\quad \langle {\mathbf Y}_A \rangle_i =\langle {\mathbf Y}_A \rangle_\Psi
\label{cond1}
\end{equation}

But eqn.(\ref{subparallel}) is precisely the requirement that all the $|a_i\rangle_A$ are also minimum
uncertainty states for ${\mathbf X}_A,{\mathbf Y}_A$. In addition, the second of the condition in 
eqn.(\ref{extraschwarz}) must be individually satisfied,
which means $(\Delta X_A)_i^2/(\Delta Y_A)_i^2$ should be the same for
all $i=1\dots s$. These constitute a
generalization of conditions (i)-(iv) spelt out earlier.

Therefore, eqn.(\ref{subparallel}) is the key to whether entangled states can saturate the equality
in the uncertainty relations (see also \onlinecite{englert},\onlinecite{englertbook}). 
What this equation means is that in the subspace spanned by $|a_i\rangle_A$, the operators
${\tilde{\mathbf X}}_A,{\tilde{\mathbf Y}}_A$ are zero. It is instructive to list a few possibilities at this
stage:(a) the operator ${\mathbf R}_A={\tilde{\mathbf X}}_A+\Gamma~{\tilde{\mathbf Y}}_A$ does not have any degenerate 
eigenfunctions. In this case entangled states can not saturate the equality;(b) ${\mathbf R}_A$ has degenerate
eigenstates but they also happen to be \emph{simultaneous} eigenstates of both ${\mathbf X}_A,{\mathbf Y}_A$.
In this case the equality will be satisfied in a trivial way in the sense that all uncertainties vanish in $|\Psi\rangle$s. 


Now, if the bipartite entangled state is such that $s = d_A$, the subspace in which the operators ${\tilde{\mathbf X}}_A,{\tilde{\mathbf Y}}_A$ becomes the entire Hilbert space ${\cal H}_A$ and this will be a realisation of case (b)
above.
Now for qubits, the Hilbert space is 2-dimensional which is equal to the
minimal Schmidt rank 2, required for a state to be entangled. Thus, our
result implies that for qubits, the lower bound of HUR or SR cannot be
attained, if the state is entangled.



Therefore, for $s=d_A$, which is the maximum possible value for s, the equality for entangled states can only be
realised trivially. On the other hand,
if $s < d_A$, the above argument does not hold, and minimum uncertainty
equality can be attained, as exemplified by Englert.\cite{englert}
The HUR for mixed states, which is related to the problem of uncertainty in
entangled states, has been studied before,\cite{luo} but that study
doesn't address the issue of minimum uncertainty.

\subsection{States of fixed angular momentum}

Now we consider the finite dimensional Hilbert space of $d_A=2j+1$, spanned by angular momentum states
with fixed value of ${\mathbf J}^2 = j(j+1)$. We only consider the case where the operators are linear
combinations of ${\mathbf J}_i$.
 The minimum uncertainty states 
in this case can be taken, without any loss of generality, to be
the eigenstates $|j,j\rangle,
|j,-j\rangle$ of ${\mathbf J}_z$.\cite{peres} For both these states, as already noted before, $\langle {\mathbf J}_x \rangle
= \langle {\mathbf J}_y \rangle = 0$. Eqn.(\ref{subparallel}) reads, in this case
\begin{equation}
\label{Jsubparallel}
\{{\mathbf J}_x+\Gamma {\mathbf J}_y\}|j,\pm j\rangle = 0
\end{equation}
Decomposing ${\mathbf J}_x+\Gamma{\mathbf J}_y$ as
\begin{equation}
{\mathbf J}_x+\Gamma{\mathbf J}_y = \frac{1-i\Gamma}{2}{\mathbf J}_++\frac{1+i\Gamma}{2}{\mathbf J}_-
\end{equation}
where ${\mathbf J}_\pm$ are the angular momentum ladder operators, and recalling
\begin{equation}
{\mathbf J}_\pm |j,m\rangle = \sqrt{(j\mp m)(j\pm m +1)}|j,m\pm 1\rangle
\end{equation}
it can easily be seen that both equations of eqn.(\ref{Jsubparallel}) can not be simultaneously satisfied.
Specifically, $|j,j\rangle$ solves it for $\Gamma = i$, and $|j,-j\rangle$ satisfies it with $\Gamma = -i$.
This proves that for the system under consideration no entangled state
saturates either the HUR or SR equality, for ${\mathbf J}_x,~{\mathbf J}_y$.
However, one can have other
observables for which the lower bound in the uncertainty relation can
be achieved.\cite{englert}




\subsection{Infinite Dimensional Hilbert Spaces}

When the Hilbert spaces ${\cal H}_A,{\cal H}_B$ are {\em infinite dimensional}, 
the whole analysis needs to be done carefully as the Schmidt decomposition
for continuous variables has many nuances. In general, states which are
entangled in discrete variables can be Schmidt decomposed. So, an analysis
similar to the one in section 3.1 is expected to go through. There has been
some recent work which shows that states with continuous variables can also
be Schmidt decomposed with discrete sets of orthogonal functions.\cite{lamata}

Let us consider the class of entangled states which can be decomposed in
the form
\begin{equation}
|\Psi\rangle = \sum_{i=1}^{\infty}~c_i|a_i\rangle_A~|b_i\rangle_B
\label{schmidt_inf}
\end{equation}
where $|a_i\rangle_A,|b_i\rangle_B$ are orthonormal basis vectors in
${\cal H}_A,{\cal H}_B$ respectively.
In such a case, our initial arguments for the finite-dimensional case will
also go through here, and for any two observables ${\mathbf X}$ and
${\mathbf Y}$, we arrive at the equation
\begin{equation}
\label{parallel_inf}
\{(X_A-\langle X_A \rangle_\Psi)+\Gamma~(Y_A-\langle Y_A \rangle_\Psi)\}
|a_i\rangle_A = 0
\end{equation}
for every i. Taking the inner product of this equation with $|a_i\rangle_A$, one gets:
\begin{equation}
(\langle X_A \rangle_i - \langle X_A \rangle_\Psi) +\Gamma~(\langle Y_A \rangle_i-\langle Y_A \rangle_\Psi)=0
\end{equation}
for every i. Here $\langle O_A \rangle_i$ is the expectation value of $O_A$ in $|a_i\rangle_A$.
But due to the 
imaginary nature of $\Gamma$, and real nature of all the expectation values, this is possible if and only if
\begin{equation}
\langle X_A \rangle _i = \langle X_A \rangle_\Psi~~\quad\quad\quad \langle Y_A \rangle_i =\langle Y_A \rangle_\Psi
\label{expec_inf}
\end{equation}
Similar manipulations on (\ref{parallel_inf}) lead us to the following results,
\begin{equation}
(\Delta X_A)_i^2 (\Delta Y_A)_i^2 =  {1\over 4}|\langle [X_A,Y_A] \rangle_i|^2
\label{uncmin_inf}
\end{equation}
\begin{equation}
(\Delta X_A)_i^2/(\Delta Y_A)_i^2 =  -2\Gamma^2 .
\label{uncratio_inf}
\end{equation}
Eqns (\ref{expec_inf},\ref{uncmin_inf},\ref{uncratio_inf}) constitute a
generalization of conditions (i)-(iv) spelt out earlier.

Let us now ask when will these conditions be satisfied.
Multiplying (\ref{parallel_inf}) by $\langle a_j|$ ($j \ne i$), we get
\begin{equation}
\langle a_j|{\mathbf X_A}|a_i \rangle = \Gamma~
\langle a_j|{\mathbf Y_A}|a_i \rangle
\label{offdiag1}
\end{equation}
Taking complex conjugate of both sides, and using Hermiticity of $X_A, Y_A$,
we get
\begin{equation}
\langle a_i|{\mathbf X_A}|a_j \rangle = \Gamma^{*}
\langle a_i|{\mathbf Y_A}|a_j \rangle
\end{equation}
Since the above is true for all i,j, we can write
\begin{equation}
\langle a_j|{\mathbf X_A}|a_i \rangle = \Gamma^{*}
\langle a_j|{\mathbf Y_A}|a_i \rangle
\label{offdiag2}
\end{equation}
Eqns (\ref{offdiag1}) and(\ref{offdiag2}) imply that
$\langle a_i|{\mathbf X_A}|a_j \rangle = 0$ and
$\langle a_i|{\mathbf Y_A}|a_j \rangle = 0$, for all i, j ($i \ne j$),
which in turn means that both ${\mathbf X_A}$ and ${\mathbf Y_A}$ are
diagonal in the subspace of states $\{|a_i\rangle\}$ for $i=1,s$.
Englert\cite{englert} has given an ingeneous example, which
falls in the class described above, to show that certain entangled
states can indeed nontrivialy saturate the equality for carefully chosen
operators.

However, there are states for which $s$ is infinite in such a way that the
sum over $i$ in (\ref{schmidt_inf}) goes over {\em all} the states of the
orthonormal set $\{|a_i\rangle\}$, and all $c_i$ are nonzero. For such states 
(\ref{offdiag1}) and (\ref{offdiag2}) imply that the minimum uncertainty
equality can be attained only when all $|a_i\rangle$ are simulatneous,
{\em degenerate} eigenstates of ${\mathbf X_A}$ and ${\mathbf Y_A}$.
That cannot happen for non-commuting observables. So, for such states
{\em no two non-commuting variables can attain minimum uncertainty}.
Lot of interesting entangled states, in infinite dimensions, fall in
this class, and no two observables will be able to saturate
the equality in HUR. Variants of the generalized form of EPR state
(\ref{newstate}) fall in this category.\cite{kim} From a practical point
of view, the states of entangled photons produced in parametric
down-conversion, also fall in this class.\cite{lamata}

Extreme care must however be exercised in handling infinite dimensional
Hilbert spaces.  A well known example is the issue of cyclicity of
traces. In finite dimensional Hilbert spaces, for any two operators
$tr\,{\mathbf A}{\mathbf B}=tr\,{\mathbf B}{\mathbf A}$. In infinite
dimensional case this is not always true, the prime example being
when the operators in question are ${\mathbf P},{\mathbf Q}$.  A naive
analysis based on a discrete, but countably infinite basis, could easily
miss this. In the particular example of the cyclicity of traces, what
causes the problem is that certain sums over the discrete indices fail to
converge. So, in our analysis, it is tacitly assumed that all relevant
expectation values exist. Another such subtlety, as was pointed out by
von Neumann long ago, is that in the infinite dimensional case it is
not always true that every self-adjoint operator can be brought to the
diagonal form.\cite{neumann2}

In view of the extreme importance of these concepts, we briefly clarify
their meaning in contemporary terms. Given an operator $\mathbf{L}$ on a
Hilbert space ${\cal H}$, the \emph{domain} of $\mathbf{L}$ is the set
of all vectors ${\rm u}$ in ${\cal H}$ such that $\mathbf{L}{\rm u}$
is well-defined. The operator L is said to be \emph{Hermitian} if it
satisfies $(u,\mathbf{L}v)=(\mathbf{L}u,v)$, for all u,v belonging to the
domain $D_L$ of $\mathbf{L}$, and  where $(u,v)$ denotes the \emph{inner
product} of the vectors u,v. The \emph{adjoint} of $\mathbf{L}$ is
defined as the operator $\mathbf{L}^\dag$ such that $(\mathbf{L}^\dag
u,v)=(u,\mathbf{L}v)$ for every u in $D_{L^\dag}$ and every v in $D_L$. It
should be carefully noted that at this stage nothing is said about the
domain $D_{L^\dag}$ of $\mathbf{L}^\dag$.  It need not be the same as
$D_L$. A \emph{self-adjoint} operator L must satisfy the {\bf twin}
requirements: i. $\mathbf{L}=\mathbf{L}^\dag$, and ii. $D_L=D_{L^\dag}$.

As the second condition generally does not hold for Hermitian operators,
it follows that every self-adjoint operator is Hermitian, but the converse
is not true. Every \emph{bounded} self-adjoint operator admits the so
called \emph{spectral theorem} over the \emph{entire} Hilbert space.
On the other hand, an \emph{unbounded} self-adjoint operator admits a
spectral theorem only over its domain. If \emph{diagonalization} or
more precisely the ability to bring an operator into diagonal form,
is understood to be equivalent to a spectral theorem over the whole
Hilbert space, unbounded self-adjoint operators can not be brought to
diagonal form over the entire Hilbert space.  An authoritative source
on these matters are the books by Reed and Simon.\cite{reedsimon} All
this rigour notwithstanding, physicists continue to use diagonal forms
even for unbounded operators like $\mathbf{P,Q}$, following Dirac and
his use of \emph{delta functions} which von Neumann  calls as ``improper
functions".

\subsection{The case of P and Q}

The case when ${\mathbf X},{\mathbf Y}$ are the momentum ${\mathbf P}$ and 
position ${\mathbf Q}$ operators, is a special one, and we treat it separately
here.
There are two possibilities: (i) both ${\cal H}_A,{\cal H}_B$ are infinite dimensional,
or (ii) only ${\cal H}_A$ is infinite dimensional. In both cases one gets the analog of (\ref{subparallel})
where now the index i runs over both continuous and discrete labels, but $|\psi_{a_i}\rangle$ need not be mutually orthogonal.
Since $[{\mathbf Q},{\mathbf P}] = i\hbar {\mathbf 1}$,
(\ref{subparallel}) for HUR leads, for example in
the position representation, to
\begin{equation}
(-i\hbar\frac{d}{dq}+i\Gamma_I q-(\langle {\mathbf P}\rangle_\Psi+i\Gamma_I \langle {\mathbf Q} \rangle_\Psi)\psi_{a_i}(q) = 0.
\end{equation}
The solution is
\begin{equation}
\psi_{a_i}(q) = C e^{i\langle P\rangle_{\Psi}q/\hbar}
\exp\left(-{|\Gamma|(q-\langle Q\rangle_{\Psi})^2\over 2}\right) .
\end{equation}
The implication is that HUR equality can be acheived only when all the
$\psi_{a_i}(q)$ are {\em same} Gaussian states 
centered around $\langle {\mathbf Q} \rangle_{\Psi}$, momentum centered around $\langle {\mathbf P}\rangle_{\Psi}$ and with width $\Delta Q =\frac{\hbar}{\Gamma_I}$. Thus $|\Psi\rangle$ can not be entangled. For the SR case, since $\Gamma$ has both real and imaginary parts, the
minimum uncertainty states acquire an additional phase $e^{i\Gamma_R q^2/2}$, but the corresponding $|\Psi\rangle$ is still
disentangled.

The result of our analysis is that the minimum uncertainty
equality for P and Q cannot be achieved for {\em any} entangled state.
This result for P and Q, however, is not very novel, as it is a straightforward
consequence for some earlier more general results on mixed state uncertainty
relation for P and Q.\cite{bastiaans1,bastiaans2,bastiaans3,manko}

\subsection{Multipartite Entanglement}
The general analysis for the bipartite case is enough to address the same issue for 
multipartite case also. The crucial issue is whether eqn.(\ref{subparallel}) admits
degenerate solutions or not. If it does, the answer in both the bipartite
and multipartite cases is the same, namely, entangled states can
saturate the equality.  This is so as one can build entangled states,
bipartite or multipartite, with these distinct states. On the other
hand, if eqn.(\ref{subparallel}) has only one solution, neither in the
bipartite case nor in the multipartite case can entangled states saturate
the equality.

\section{The Mixed State Version}
The question we have raised about entanglement and minimum uncertainty can equivalently be formulated
as a problem about \emph{mixed} states and minimum uncertainty of a single quantum system. In this
section we discuss some generalities and previously known results, and our own variational
formulation that elegantly unifies the HUR and SR cases.

 Since we have considered operators 
acting only on one of the subspaces, 
expectation values in the
entangled pure states of the larger system
are equivalent to expectations taken in a \emph{mixed} state
of the subsystem. To see this, consider a generic pure entangled state of a bipartite system:
\begin{equation}
\label{eq:entanglegen}
|\Psi\rangle = \sum_{ij}\,d_{ij}|\alpha_i\rangle_A|\beta_j\rangle_B\quad\quad\sum_{ij}\,|d_{ij}|^2=1
\end{equation}
The expectation values in such a state of operators of the type ${\mathbf O}_A\otimes{\mathbf 1}_B$
are given by
\begin{equation}
\label{eq:entangledmixed}
\langle \Psi|{\mathbf O}_A\otimes{\mathbf 1}_B|\Psi\rangle = \sum_{ik}(\sum_j d_{ij}d^*_{kj})O^A_{ki} = tr\: {\mathbf\rho}^A\,{\mathbf O}_A
\end{equation}
where $\rho^A_{ik} =\sum_j\,d_{ij}d^*_{kj}$ is the mixed density matrix for system A.
Though we have shown this for finite dimensional systems, it is expected to be valid even for infinite dimensional cases.
 
Uncertainty relations can obviously be formed for mixed states too as they only involve various
expectation values, which are well defined both for pure as well as mixed states. Let us see how the simpler HUR works.
 Consider a state ${\mathbf \rho}$, and two non-commuting
operators ${\mathbf X},{\mathbf Y}$. The expectation values of these operators in the given state are $tr {\mathbf \rho}{\mathbf X}$
and $tr {\mathbf \rho}{\mathbf Y}$ respectively. As before, introduce
 ${\tilde{\mathbf X}} = {\mathbf X}
-tr {\mathbf \rho}{\mathbf X}$, and likewise for ${\tilde{\mathbf Y}}$.

 The analog of the Schwarz inequality for mixed states is to consider the positive semi-definite quantity, with $a$  real ,
\begin{equation}
\label{eq:dhur}
D_{hur}(a) = tr {\mathbf \rho}\,({\tilde{\mathbf X}}+ia{\tilde{\mathbf Y}})^\dag
\,({\tilde{\mathbf X}}+ia{\tilde{\mathbf Y}})
\ge 0
\end{equation}
 Writing out explicitly
\begin{equation}
\label{eq:dhur2}
D_{hur}(a) = (\Delta X)_\rho^2 +a^2(\Delta Y)_\rho^2 + ia \langle[{\mathbf X},
{\mathbf Y}]\rangle_\rho
\end{equation}
The discriminant condition for the positive semi-definiteness of $D_{hur}(a)$ can be shown to be precisely
the same as the HUR. But we formulate this somewhat differently, through a variational approach, as that
will also provide a natural means for addressing the question of states saturating the bound.

For given ${\mathbf {\rho, X, Y}}$, $D(a)$ can be considered as a positive
semi-definite function.  Clearly, its minimum exists as
\begin{equation}
\label{eq:dhurmin}
\frac{d^2}{da^2}\, D_{hur}(a) = (\Delta Y)^2_\rho > 0
\end{equation}
and the minimum occurs at
\begin{equation}
\label{eq:dhuramin}
a_{min} = -\frac{1}{2}\,\frac{\langle i[{\mathbf X},{\mathbf Y}]\rangle_\rho}{(\Delta Y)_\rho^2}
\end{equation}
Consequently, the minimum value of $D_{hur}(a)$ can be calculated to be
\begin{equation}
\label{eq:dhurminfin}
D^{min}_{hur} = (\Delta X)_\rho^2 - \frac{1}{4}\,\frac{|\langle i[{\mathbf X},
{\mathbf Y}]\rangle_\rho|^2}{(\Delta Y)_\rho^2}\,\ge 0
\end{equation}
 This is nothing but the HUR inequality. 

Now we address the issue of determining states, if any, that would saturate the inequality in HUR. After explaining
our method, we shall make a comparison with the approach of Englert,\cite{englertbook} who has also addressed the same issue. Notice that
 the last equation can be rewritten as
\begin{equation}
\label{eq:Chur}
D^{min}_{hur} = tr {\mathbf \rho}\,C_{hur}^\dag\,C_{hur}\,\ge 0
\end{equation}
 where
\begin{equation}
\label{eq:Chur2}
C_{hur} = {\tilde{\mathbf X}} + ia_{min}\,{\tilde{\mathbf Y}}
\end{equation}
We  write this out explicitly by using eqn.(\ref{eq:dhuramin}),
\begin{equation}
\label{eq:Chur3}
C_{hur} = {\tilde{\mathbf X}} + \frac{\langle[{\mathbf X},{\mathbf Y}]\rangle_\rho}{2(\Delta Y)_\rho^2}\,{\tilde{\mathbf Y}}
\end{equation}
 The minimum uncertainty states, denoted by ${\mathbf \rho}_{min}$ must satisfy
\begin{equation}
\label{eq:hurrhomin}
tr {\mathbf \rho}_{min}\,C_{hur}^\dag\,C_{hur} = 0
\end{equation}
 Before showing how to solve this, we shall first establish a similar result for the SR case. In the light of our earlier
treatment of the SR case, we introduce the positive semi-definite operator
$D_{sr}(\Gamma)$, with $\Gamma$ a \emph{complex} quantity:
\begin{equation}
\label{eq:dsr}
D_{sr}(\Gamma) = tr {\mathbf \rho}\,({\tilde{\mathbf X}}+\Gamma{\tilde{\mathbf Y}})^\dag
\,({\tilde{\mathbf X}}+\Gamma{\tilde{\mathbf Y}})
\ge 0
\end{equation}
 Writing out explicitly
\begin{equation}
\label{eq:dsr2}
D_{sr}(\Gamma) = (\Delta X)_\rho^2 +|\Gamma|^2(\Delta Y)_\rho^2 
+ i\Gamma_I \langle[{\mathbf X}, {\mathbf Y}]\rangle_\rho
+ \Gamma_R \langle\{\tilde{\mathbf X}, \tilde{\mathbf Y}\}\rangle_\rho
\end{equation}
 Now minimisation has to be done wrt both $\Gamma_R$ and $\Gamma_I$.  Clearly, the minimum exists as
\begin{equation}
\label{eq:minGamma}
\frac{d^2}{d\Gamma_R^2}\, D_{sr}(\Gamma) = 
\frac{d^2}{d\Gamma_I^2}\, D_{sr}(\Gamma) = 
(\Delta Y)^2_\rho > 0
\end{equation}
As $D_{sr}$ contains no mixed terms in $\Gamma_I, \Gamma_R$, there are no further conditions.
 The minimum occurs at
\begin{equation}
\label{eq:minGamma2}
{\Gamma_I}_{min} = -\frac{1}{2}\,\frac{\langle i[{\mathbf X},{\mathbf Y}]\rangle_\rho}{(\Delta Y)_\rho^2}
\quad\quad {\Gamma_R}_{min} = -\frac{1}{2}\,\frac{\langle \{\tilde{\mathbf X},\tilde{\mathbf Y}\}\rangle_\rho}{(\Delta Y)_\rho^2}
\end{equation}
 Consequently, the minimum value of $D_{sr}(\Gamma)$ is
\begin{equation}
\label{eq:}
D^{min}_{sr} = (\Delta X)_\rho^2 - \frac{1}{4}\,\frac{|\langle i[{\mathbf X},
{\mathbf Y}]\rangle_\rho|^2
+|\langle\{{\tilde{\mathbf X}},{\tilde{\mathbf Y}}\}\rangle_\rho|^2}{(\Delta Y)_\rho^2}\,\ge 0
\end{equation}
 This is just the SR inequality. 

It is clear that $D^{min}_{sr}$ can be rewritten in the useful form
\begin{equation}
\label{eq:dsrcsr}
D^{min}_{sr} = tr {\mathbf \rho}\,C_{sr}^\dag\,C_{sr}\,\ge 0
\end{equation}
 where
\begin{equation}
\label{eq:csr}
C_{sr} = {\tilde{\mathbf X}} + \Gamma_{min}\,{\tilde{\mathbf Y}}
\end{equation}
 Writing out explicitly
\begin{equation}
\label{eq:csr2}
C_{sr} = {\tilde{\mathbf X}} - \frac{\langle i[{\mathbf X},{\mathbf Y}]\rangle_\rho
+\langle\{{\tilde{\mathbf X}},{\tilde{\mathbf Y}}\}\rangle_\rho}{2(\Delta Y)_\rho^2}\,{\tilde{\mathbf Y}}
\end{equation}
 The states that saturate the bound, denoted by ${\mathbf \rho}^{SR}_{min}$ must satisfy
\begin{equation}
\label{eq:srrhomin}
tr {\mathbf \rho}^{SR}_{min}\,C_{sr}^\dag\,C_{sr} = 0
\end{equation}
This is of the same form as eqn.(\ref{eq:hurrhomin}). Consequently the same techniques, to be described shortly,
can be used to solve both of them.
\subsection{Comparison with Englert}
Englert\cite{englertbook} has already addressed the problem of determining the mixed states that saturate the uncertainty
bound. He has only considered the HUR case. He introduces an operator $C$, similar in spirit to our $C_{hur}, C_{sr}$, but
with some important differences. 
 The minimum uncertainty conditions for the Englert operators are
$\langle C^\dag\,C\rangle =0$ if the sign of $\langle i[{\mathbf X},{\mathbf Y}]\rangle = -1$,
and $\langle C\,C^\dag \rangle =0$ otherwise. 
 So, in his approach, one has to explicitly keep track of these
possibilities. Using the explicit forms of his operators, it is easy to see that 
in the former case $C = C_{hur}$ whereas in the latter case $C^\dag = C_{hur}$.
Thus, though the minimum uncertainty conditions in terms of his C-operator seem to depend on the context, in terms of our $C_{hur}$,
in both cases the minimum uncertainty condition is the \emph{same}.
 In the SR case, signs of both $\langle i[{\mathbf X},{\mathbf Y}]\rangle$ and $\langle \{{\tilde{\mathbf X}},{\tilde{\mathbf Y}}\}\rangle$ 
have to be kept track of had we followed Englert's method. That would have been four distinct cases for analysis. On the other hand, 
in terms of our $C_{sr}$, only one condition suffices exactly
as in the HUR case.
\subsection{Mixed State Analysis}
 So far our analysis of the density matrix approach has been quite general. Though our objective is to address the
issue of saturation of the HUR and SR bounds for mixed states, we did not specifically use the mixed nature of the
states till now. Now we specialize to mixed
states, and ask: \emph{can mixed states saturate the uncertainty bounds?} Let such a mixed state  $\rho_{min}$ be represented by
\begin{equation}
\label{eq:}
\rho_{min} =\sum\,p_i\,\rho_i\quad\quad p_i\ge 0
\end{equation}
The $\rho_i$ above are \emph{pure} state density matrices, of rank 1. The minimum uncertainty condition in both the HUR and SR cases is of the form
\begin{equation}
\label{eq:}
tr \rho_{min}\,C^\dag\,C =0=\sum_i\,p_i\,tr\,\rho_i\,C^\dag C=0
\end{equation}
 This implies, for all those $i$ for which $p_i\,\ne 0$, 
\begin{equation}
\label{eq:}
tr\,\rho_i\,C^\dag\,C =0
\end{equation}
 Representing $\rho_i$ as $|a_i\rangle\langle a_i|$,
one gets
\begin{equation}
\label{eq:}
C|a_i\rangle = 0
\end{equation}
 This is exactly the same key equation we got earlier.
\subsection{Known equivalent results for mixed states}
We now enumerate known results for mixed states that has a bearing on our
original conjecture. The earliest known result that is of relevance is
due to Bastiaans.\cite{bastiaans1,bastiaans2,bastiaans3} He formulated
these in the context of partially coherent light. In our notation,
his main result can be
expressed as
\begin{equation}
\label{eq:bastiaans}
\Delta P \Delta Q > \frac{\hbar}{2}\cdot\frac{8}{9\mu}
\end{equation}
where $\mu = tr \rho^2$ is the so called \emph{purity} of a state, and for mixed states it is always less
than unity. Thus the result of Bastiaans states that no mixed state of purity $\mu < \frac{8}{9}$ can ever saturate the HUR bound, and
because of the equivalence of mixed state expectations to expectations in pure entangled states, this
is the same as the result that for P,Q, no entangled state with an equivalent purity of less than $\frac{8}{9}$ can saturate the HUR bound. 
This is of course weaker than our result that for these observables \emph{no} entangled state can saturate the bound. Additionally,
Bastiaans did not address the SR case.

In a very thorough and scholarly work on the uncertainty relations for mixed states, Dodonov and Man'ko\cite{manko}
have addressed both these aspects. Firstly, they showed that the abovementioned result of Bastiaans was valid only for mixed states of
very low purity. They obtained the following result for mixed states of \emph{arbitrary} purity:
\begin{equation}
\label{eq:mankododonov}
(\Delta P)^2(\Delta Q)^2 -\frac{1}{4}\langle \{\tilde{\mathbf P},\tilde{\mathbf Q}\}\rangle^2 \ge 
\frac{\hbar^2}{4}\Phi^2(\mu)
\end{equation}
where $\Phi(\mu)$ is a function that is \emph{always} greater than unity for mixed states i.e $\mu < 1$. 
An approximate(to within 1\%) analytical form of $\Phi(\mu)$
being
\begin{equation}
\label{eq:mankophi}
\Phi(\mu) = \frac{4+\sqrt{16+9\mu^2}}{9\mu}
\end{equation}
Thus our result for P,Q is completely equivalent to this earlier result. In a recent very interesting work Man'ko and 
collaborators\cite{mankoexp} have even experimentally verified eqn.(\ref{eq:mankododonov}). 

Another characterisation of mixed states is through their \emph{entropy}
(see \onlinecite{neumann1} and \onlinecite{shannon}). 
Dodonov and Man'ko also gave the following
very interesting form of the mixed state uncertainty relations:
\begin{equation}
\label{eq:mankoentropic}
(\Delta P)^2(\Delta Q)^2 -\frac{1}{4}\langle \{\tilde{\mathbf P},\tilde{\mathbf Q}\}\rangle^2 \ge 
\frac{\hbar^2}{4}\:\left(1+\frac{2}{e^\beta-1}\right)^2
\end{equation}
where the parameter $\beta$ is related to the entropy, $S = -tr\,\rho\ln\rho$, of the mixed state as follows:
\begin{equation}
\label{eq:mankoStobeta}
S =\beta(e^\beta-1)^{-1}-\ln (1-e^{-\beta}) 
\end{equation} 
It again follows that for P,Q no mixed state, for which $S\ne 0$, can the uncertainty bound for HUR and SR be saturated.

But we wish to emphasize that the techniques used by Dodonov and Man'ko can not be easily generalized to arbitrary observables,
while our proofs easily extend to those cases.

Based on his methods, Englert gave counterexamples\cite{englert} to our original claim, both for finite dimensional and infinite dimensional
cases. While his counterexamples definitively showed our conjecture to be not generally valid, they left open the interesting,
and useful,
question of finding systems and observables for which the conjecture \emph{is} true. That is the main purpose of this paper.

\section{Conclusion}

In conclusion, we have shown that entanglement puts a bound on the product
of uncertainties of non-commuting observables, for certain class of 
systems and states. 
For finite-dimensional Hilbert spaces we have show that no two observables 
can attain the minimum uncertainty equality if the Schmidt rank of the
entangled state is maximal. Since the Schmidt rank of any entangled state, in
a 2-dimensional Hilbert space, is always maximal, one of the consequences is
that no two observables of a {\em qubit} can attain minimum uncertainty if
the state is entangled. For infinite-dimensional Hilbert spaces, we have
shown that there is a class of physically significant states for which
no two observables can satisfy the minimum uncertainty equality.
The problem for the case of HUR was equivalently addressed for
mixed single-particle states by \cite{englert,englertbook}. Mention must also
be made of the so called \emph{entropic} uncertainty relations.\cite{everettorig,everett,deutsch,mds,mdsbook,rudnicki} Brillouin appears
to be the first one to think of information theoretic approach to the uncertainty relations.\cite{brillouin} As stressed first by
Everett,\cite{everettorig} these relations are stronger than HUR.
But the issue of states saturating
the equality in the entropic and other forms of uncertainty relations is a complex one,
and the immediate connection to our results is not obvious. 

\section*{Acknowledgments}

Tabish Qureshi thanks Pankaj Sharan
and Deepak Kumar for useful discussions. Hari Dass thanks M.D. Srinivas for
illuminating discussions. He also acknowledges support from
Department of Science and Technology to the project IR/S2/PU-001/2008.  
Aditi Sheel thanks the Centre for Theoretical Physics for providing her
the hospitality of the Centre during the course of this work.

\end{document}